\documentclass[amsmath,aps,prc,showpacs,reprint,superscriptaddress,floatfix,nofootinbib]{revtex4-1}
\usepackage{graphicx}
\usepackage{float}
\usepackage[caption = false]{subfig}
\usepackage{dcolumn}
\usepackage{bm}
\usepackage{longtable}
\usepackage{afterpage}
\usepackage[colorlinks=true, citecolor=blue, linkcolor = blue, urlcolor=blue]{hyperref}

\begin{document}
\title{High-spin spectroscopy in \textsuperscript{207}At: Evidence of a 29/2$^{+}$ isomeric state}
\author{Khamosh Yadav}
\author{A. Y. Deo}
\email{Corresponding author: ajay.deo@ph.iitr.ac.in}
\author{Madhu}
\author {Dhananjaya Sahoo}
\author{P. C. Srivastava}
\affiliation{Department of Physics, Indian Institute of Technology Roorkee, Roorkee 247667, India}
\author{Saket Suman}
\author{S. K. Tandel}
\affiliation{School of Physical Sciences, UM-DAE Centre for Excellence in Basic Sciences, University of Mumbai, Mumbai 400098, India}
\author{A. Sharma}
\affiliation{Department of Physics, Himachal Pradesh University, Shimla 171005, India}
\author{I. Ahmed}
\author{K. Katre}
\author{K. Rojeeta Devi}
\author{Sunil Dutt}
\author{Sushil Kumar}
\author{Yashraj}
\author{S. Muralithar}
\author{R. P. Singh}
\affiliation{Inter-University Accelerator Centre, Aruna Asaf Ali Marg, New Delhi 110067, India}
\date{\today}
\begin{abstract}
Yrast and near-yrast states above the known 25/2$^{+}$ isomer in $^{207}$At
are established for the first time. The level scheme is extended up to
47/2$\hbar$ and 6.5 MeV with the addition of about 60 new $\gamma$-ray
transitions. The half-life of the 25/2$^{+}$ isomer is revisited and a
value of $T_{1/2}$ = 107.5(9) ns is deduced. Evidence of a hitherto unobserved
29/2$^{+}$ isomer in $^{207}$At is presented. A systematic study of $B(E3)$ values
for the transitions de-exciting the 29/2$^{+}$ isomer in the neighboring odd-$A$ At
isotopes suggests a half-life in the 2$-$4.5 $\mu$s range for this state in $^{207}$At.
The experimental results are compared with large-scale shell-model
calculations performed using the KHM3Y effective interaction in the $Z$ = 50$-$126,
$N$ = 82$-$184 model space and an overall good agreement is noted between the theory
and the experiment. A qualitative comparison of the excited states and the isomers
with analogous states in neighboring nuclei provides further insight into
the structure of $^{207}$At.
\end{abstract}
\maketitle

\section{Introduction}\label{sec:I}
Neutron-deficient nuclei in the region around the doubly magic $^{208}$Pb
nucleus reveal a large variety of structural properties and phenomena
such as shape coexistence, shears bands, superdeformed bands, and various
kinds of low- and high-spin isomers \cite{Heyde,Wood,Julin,192-195Po}.
Nuclei in the vicinity of both the proton and neutron shell closures
exhibit spherical or near-spherical shapes arising due to intrinsic
degrees of freedom ({\it i.e.} particle-hole excitations) \cite{voigt}.
However, the onset of collective excitations is evident in nuclei with
several valence particles or holes outside the shell closures.
In this respect, Po and At isotopes offer a suitable ground for understanding
the shape evolution.
The level structures in the Po isotopes near the neutron shell closure
({\it N} = 126) have been interpreted in terms of intrinsic degrees of
freedom using the shell-model approach, while the collective modes of
excitation have been observed in the lighter Po isotopes \cite{206Po_1, 204_206Po}.
The Po isotopes with $N$ $\approx$ 112, viz. $^{192-195}$Po \cite{192-195Po},
exhibit oblate-deformed structures, and subsequently evolve into prolate
deformation as the $N$ = 104 neutron mid-shell is approached \cite{190Po}.
Experimentally, the shape evolution, as one moves from the spherical
to the deformed regime, is manifested in the enhancement of the $E$2 transition
strengths and the energy systematics of the first excited 2$^{+}$ and
4$^{+}$ states \cite{206Po_1,204_206Po}.
A recent study on the low-lying level structures in $^{204}$Po and
$^{206}$Po \cite{204_206Po} isotopes reports that the transition
from single-particle to collective-excitation modes with decreasing
neutron number occurs below $^{206}$Po.
With one extra proton, the level structures in odd-$A$ At isotopes
can be interpreted in terms of an unpaired proton coupled to the even-even
core of the corresponding Po isotone. Hence, the odd-$A$ At nuclei are
expected to inherit structural properties and phenomena
from the corresponding Po core, which is indeed reflected in the
earlier studies \cite{194-196At, 197At, 199At, 201_203At, 201At,
203At, 205At_1, 205At_2, 207At,209At_1,209At_2,209At_3,209At_4,211At}.
%
\begin{figure*}[ht!]
\begin{centering}
  \includegraphics*[angle = 270, width=\textwidth]{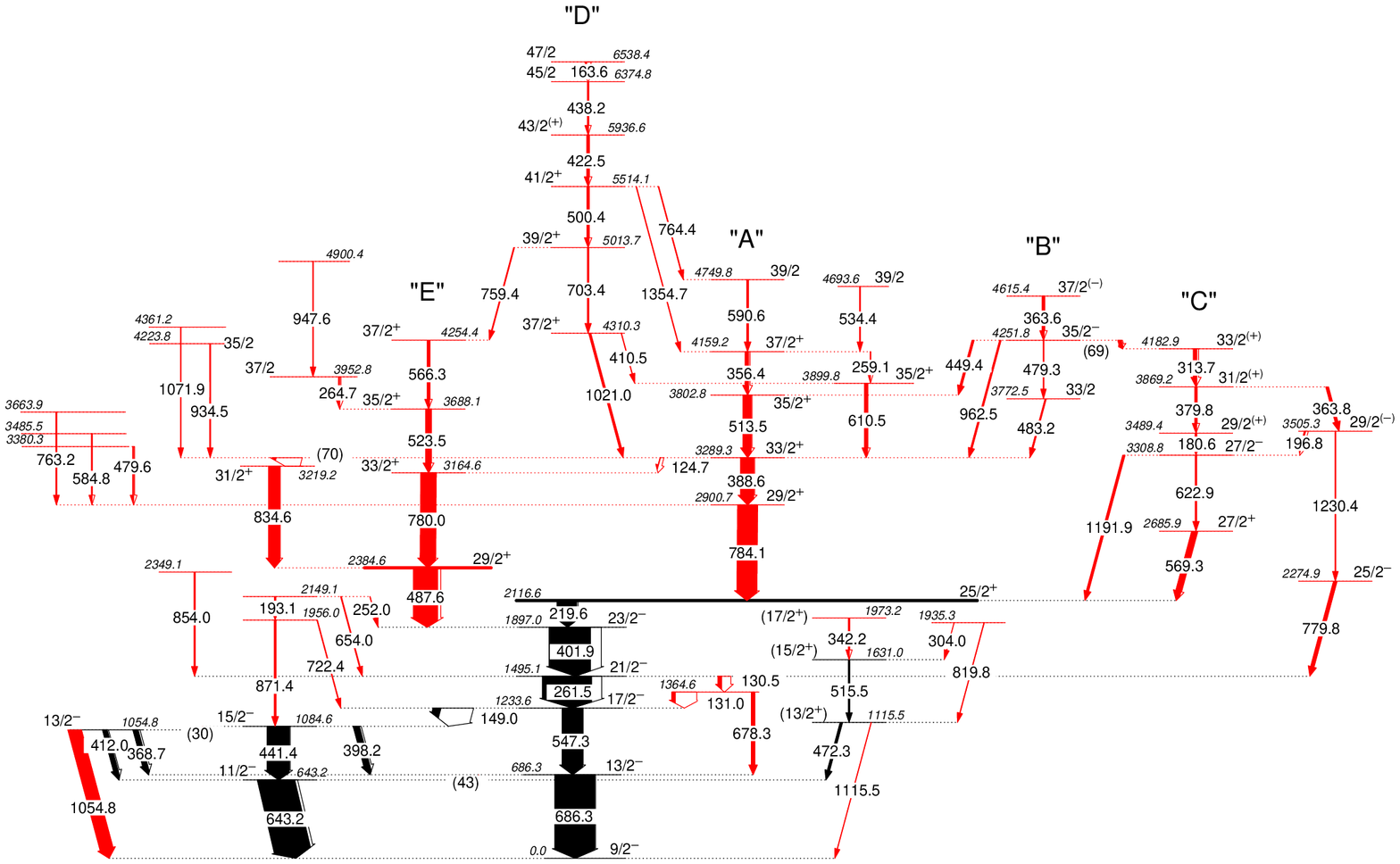}%
\end{centering}
\caption{\label{LS}The level scheme of $^{207}$At established in the present work.
The $\gamma$-ray transitions and levels are labeled with their energies in keV.
The widths of the filled and open areas of the arrows correspond to the intensity of the
$\gamma$ rays and conversion electrons, respectively. The newly identified
$\gamma$ rays and levels are presented in red color. The transitions shown in parentheses
are not observed in the present experiment, but their presence is inferred from the
coincidence relationships.
}
\end{figure*}

Several high-spin isomers have been reported in nuclei with $Z$ $>$ 82
and $N$ $\leq$ 126, which further motivates nuclear structure studies in the
above near-spherical regime. The study of nuclear isomers is pivotal as it provides
key inputs for understanding the structure of nuclei and to test the predictions
of the shell model \cite{AKJ}. Among the neutron-deficient even-$A$ Po isotopes,
isomeric 8$^{+}$, 9$^{-}$, and 11$^{-}$ states have been observed in $^{200-208}$Po
\cite{202_204_206Po, 206Po, 206Po_2, 202_204Po, 204Po, 204Po_204At, 200_202Po, Po_isotopes, 201_203Po}. 
The hindrance in the decay of the 8$^{+}$ state has been understood in terms of small
energy difference between the 6$^{+}$ and 8$^{+}$ states of the $\pi(h_{9/2}^2$) multiplet,
while the metastable nature of the 9$^{-}$ and 11$^{-}$ states is attributed to a change in
the single-particle configuration of the isomeric state and the one to which it decays
\cite{202_204_206Po, 206Po, 206Po_2, 202_204Po, 204Po, 204Po_204At, 200_202Po, Po_isotopes, 201_203Po}. 
Similarly, several high-spin isomers have also been reported in At isotopes.
An isomeric 13/2$^{+}$ level, which originates from the $\pi(i_{13/2}$)
configuration, has been observed in $^{197-203}$At isotopes \cite{199At,201At, 203At}.
In $^{197,199}$At nuclei \cite{197At, 199At}, the isomeric 13/2$^{+}$ state 
is suggested to have oblate deformation
and is populated by a strongly coupled rotational band,
while it is known to have weak oblate deformation in $^{201}$At \cite{201At}.
Also, 25/2$^{+}$ isomeric states with $T_{1/2}$ in the range of $\approx$ 14$-$108 ns
were reported in $^{203,205,207}$At \cite{203At,205At_1,205At_2,207At}.
Furthermore, an isomeric 29/2$^{+}$ state has been observed in odd-$A$ At isotopes
with $199 \leq A \leq 211$ \cite{201At,203At,205At_1,205At_2,209At_1,209At_2,209At_3},
except in $^{207}$At \cite{207At}. The systematic presence of the 29/2$^{+}$ isomeric
level in odd-$A$ At isotopes is one of the motivations for the detailed high-spin study
along with a search for the 29/2$^+$ isomer in $^{207}$At.
A comprehensive spectroscopic study in $^{207}$At ($Z$ = 85, $N$ = 122)
is also needed for understanding the shape evolution along the isotopic
chain in neutron-deficient odd-$A$ At nuclei and the effect of the proton-particle
and neutron-hole excitations.

In the previous high-spin study of $^{207}$At \cite{207At},
excited states were investigated using the $^{204}$Pb($^{6}$Li, 3{\it n})$^{207}$At
reaction. The level scheme up to 25/2$\hbar$ with 2117-keV excitation energy
was established. A total of ten excited states were reported
in the study by Sjoreen {\it et al.}\cite{207At}, out of which
eight levels were established with firm spin-parity.
It is interesting to note that the isomeric 25/2$^{+}$ level
at 2117 keV is the only positive-parity state known in the level
scheme of $^{207}$At \cite{207At}.
On the other hand, the level schemes of the neighboring odd-$A$ At
isotopes are known to have considerable spectroscopic information
and present well-established positive- and negative-parity
sequences \cite{201At, 203At, 201_203At, 205At_2}.
In this paper, we propose an extended level scheme of $^{207}$At with
the addition of about 60 new $\gamma$-ray transitions. The excited states
above the 25/2$^{+}$ isomer are reported for the first time.
We also present the first evidence of 29/2$^{+}$ isomeric state in $^{207}$At,
which is known in all neighboring odd-$A$ At
isotopes \cite{201At,203At,205At_1,205At_2,209At_1,209At_2,209At_3}.
%
\begin{figure*}[ht!]
 \includegraphics*[width=0.9\textwidth]{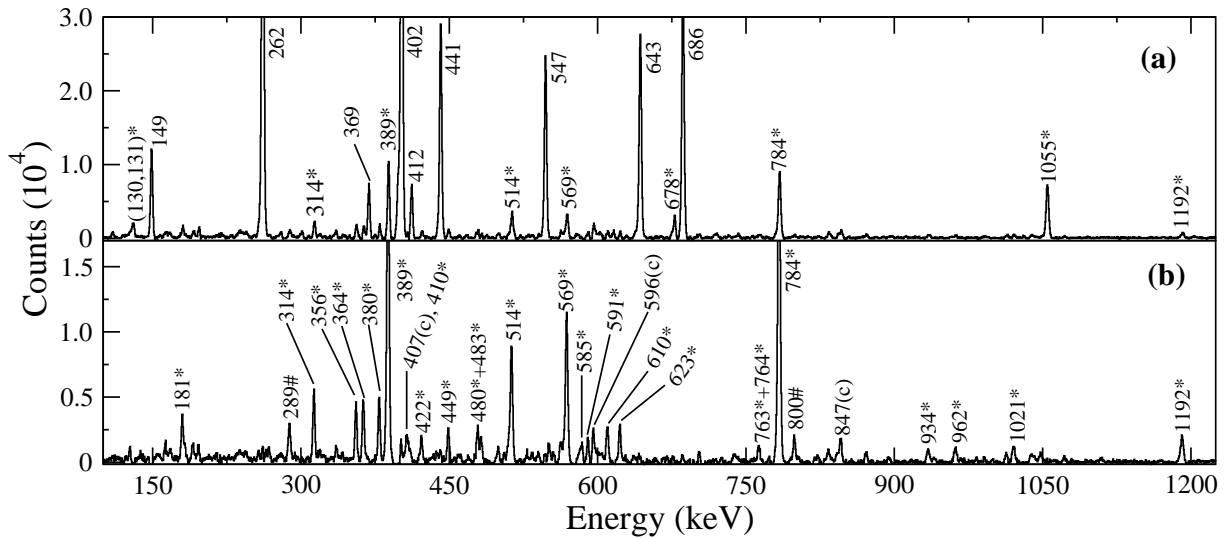}%
 \caption{\label{fig:gD_219} The $\gamma$-ray spectra illustrating the transitions
detected within (a) $\pm$ 100 ns prompt, and (b) 100$-$275 ns {\it early} coincidence windows of the
220-keV $\gamma$ ray.
The new $\gamma$ rays from $^{207}$At, which are firmly placed in the
level scheme, are marked with an asterisk.
The $\gamma$-ray transitions marked with symbol ``$\#$'' also belong to $^{207}$At.
However, their position in the level scheme could not be ascertained.
The transitions labeled with symbol (c) are contaminant lines.
}
\end{figure*}
%
%
\section{Experimental Details and Data analysis}\label{sec:II}
High-spin states in the $^{207}$At nucleus were investigated using the
$^{198}$Pt($^{14}$N, 5{\it n})$^{207}$At heavy-ion fusion-evaporation
reaction. A self-supporting $^{198}$Pt target of thickness $\approx$ 10 mg/cm$^2$
and about 92$\%$ enrichment was bombarded with a DC beam of $^{14}$N ions.
The beam with 80-, 85-, and 87-MeV energies
was delivered by the 15-UD Pelletron accelerator at Inter-University
Accelerator Centre (IUAC), New Delhi. In addition to the nucleus of interest $i.e.$ $^{207}$At,
the data also had contributions from other reaction channels, viz. $^{205,206,208}$At, and their
$\beta$-decay products. It was also observed that the data had some degree of contamination
from the Coulomb excitations of the target material and the reaction of the beam halo with the iron target frame.
The $\gamma$-multiplicity trigger was set to collect the two- and higher-fold
coincidence data, which was used to deduce the level scheme of $^{207}$At.
In addition, single-fold data were also acquired
to determine the intensity of the $\gamma$-ray transitions, whenever possible.
The Indian National Gamma Array (INGA) \cite{INGA} was utilized to record the data,
which consisted of 13 Compton suppressed clover HPGe detectors and
a Low-Energy Photon Spectrometer (LEPS) at the time of the experiment.
The detectors were located at five different angles, 32$^{\circ}$, 57$^{\circ}$,
90$^{\circ}$, 123$^{\circ}$, and 148$^{\circ}$ with respect to the beam direction.
The data were acquired using a VME-based data acquisition system \cite{VME} and written
to a disk in the ROOT \cite{ROOT} tree format. The standard $^{152}$Eu and
$^{133}$Ba sources were used to obtain the energy calibration and relative
$\gamma$-ray efficiencies of the clover Ge detectors.
The data collected at the three beam energies
were combined to generate various two- and three-dimensional histograms,
which aid in establishing the coincidence relationships,
ordering of the levels, lifetimes of the isomeric states, and the multipolarity
of the $\gamma$-ray transitions. The software packages ROOT \cite{ROOT}
and RADWARE \cite{RADWARE} were used to analyze these histograms.

The electronic time of each $\gamma$
ray was obtained from the TDC signal from the corresponding clover detector.
The coincidence windows used to construct the histograms were defined by the
time difference between the corresponding $\gamma$ rays.
In order to construct symmetric two- and three-dimensional histograms,
the $\gamma$ rays detected within 100 ns of each other were utilized.
The coincidences occurring within 100$-$200 ns
of each other were subtracted from the above histograms to account for random events.
The resulting prompt histograms were utilized to establish coincidence relationships
among the observed $\gamma$-ray transitions. Furthermore, an asymmetric $\gamma-\gamma$ matrix, termed as an
{\it early-delayed} matrix, was also constructed using the 
$\gamma$ rays detected within 100$-$275 ns of each other.
The {\it early} and {\it delayed} transitions for the above matrix are classified based on their relative electronic times.
This matrix was utilized to investigate the isomers and the correlations between the transitions across
the isomeric states. It may be noted that the useful range of the TDC was 0$-$275 ns.

The multipolarity of the $\gamma$-ray transitions were assigned on the basis
of the Directional angular Correlation from the Oriented states (DCO) ratio
measurements \cite{DCO} or intensity balance considerations. The experimental
DCO ratio ($R_\textrm{DCO}$) for a transition of interest was defined as the ratio
of its intensities in the coincidence spectra from the detectors at 148$^{\circ}$/32$^{\circ}$
to that at 90$^{\circ}$ with respect to the beam direction.
For the present detection setup, the $R_\textrm{DCO}$ value was found to
be close to 1.0 (0.5) for a stretched quadrupole (dipole) transition with a
gate on a stretched quadrupole transition, and a value of $R_\textrm{DCO}$ $\approx$
2.0 (1.0) was observed for a stretched quadrupole (dipole) transition in
a dipole gate. More details on the procedure of the DCO ratio measurements
can be found elsewhere \cite{215Fr, 216Fr}.
The parity of the excited levels was assigned on the basis of the electric or magnetic
nature of the feeding and/or depopulating $\gamma$-ray transitions.
The electromagnetic nature of the $\gamma$-ray transitions, whenever possible,
was confirmed from the polarization asymmetry ($\Delta_{asym}$) measurements \cite{IPDCO1}.
For this purpose, two asymmetric $\gamma-\gamma$ matrices were generated with one
axis comprising the energies of the $\gamma$ rays scattered in the perpendicular
or parallel crystals of the 90$^{\circ}$ detectors with respect to the reaction
plane and the other axis corresponding to the $\gamma$-ray energies detected
in the remaining detectors. Furthermore, the $\Delta_\textrm{asym}$ was defined as
\begin{equation}
 \Delta_\textrm{asym} = \frac{a(E_{\gamma})N_{\perp} - N_{\parallel}}{a(E_{\gamma})N_{\perp} + N_{\parallel}},
\end{equation}
where $N_{\perp}$ ($N_{\parallel}$) are the counts recorded in the perpendicular
(parallel) segments of the 90$^{\circ}$ detectors, where $a(E_{\gamma})$ is a geometrical
asymmetry factor which is defined as the ratio of $N_{\parallel}$ and $N_{\perp}$ for
the $\gamma$ rays from an unpolarized radioactive source. The experimental values
of $a(E_{\gamma})$ obtained using a $^{152}$Eu source were plotted as a function of
$\gamma$-ray energies and fitted with a function, $a(E_{\gamma})$ = $a_0$ + $a_1 E_{\gamma}$.
The coefficients $a_0$ and $a_1$ were found to be 1.029(3) and -7.96(363) $\times$ 10$^{-6}$, respectively.
The positive and negative values of the $\Delta_\textrm{asym}$ signify electric and
magnetic nature of the $\gamma$-ray transitions of interest, respectively.
%
%
\begin{figure}[t!]
\begin{center}
\includegraphics*[width=\columnwidth]{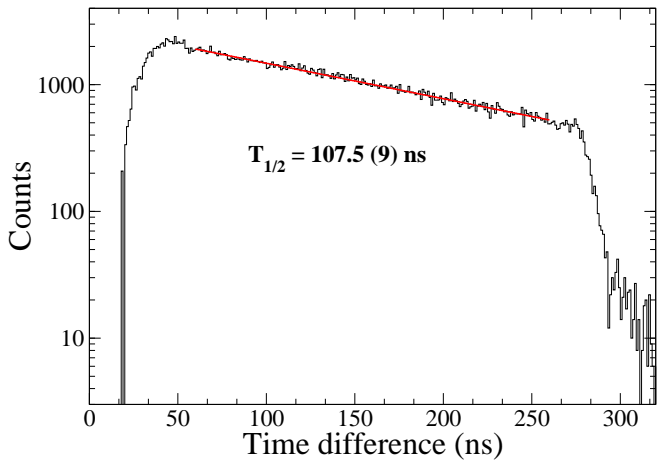}%
\caption{\label{fig:halflife} Time-distribution curve (in black color) to determine half-life of
the 25/2$^+$ state. The curve is obtained by subtracting the
electronic timing of the 784-keV transition from that of the 220-keV transition.
The contributions of the prompt time-distribution and background are removed from the curve.
The solid line (in red color) corresponds to 
an exponential fit to the time-distribution curve.
}
\end{center}
\end{figure}
%
\section{Results}\label{sec:III}
Excited states in $^{207}$At were studied earlier through $^{204}$Pb($^{6}$Li, 3$n$)$^{207}$At
heavy-ion fusion evaporation reaction \cite{207At}. This work had established
a total of ten excited levels in $^{207}$At including a 25/2$^+$ isomer with
$T_{1/2}$ = 108(2) ns at 2117 keV \cite{207At}. An extended level scheme of
$^{207}$At deduced from the present study is shown in Fig. \ref{LS}.
The level scheme below the 25/2$^{+}$ isomer \cite{207At} is revisited
and found to be consistent with the present data. Figure \ref{fig:gD_219}(a)
illustrates the $\gamma$-ray transitions in coincidence with the 220-keV $\gamma$ ray
which de-excites the 25/2$^{+}$ isomer. In addition to the known transitions,
several new transitions were observed in the above coincidence spectrum.
Figure \ref{fig:gD_219}(b) presents $\gamma$-ray transitions which precede the 220-keV
$\gamma$ ray within a 100$-$275 ns time window.
The relative enhancement in the intensity of the 314-, 389-, 514-, 569-, 784-,
and 1192-keV $\gamma$ rays in Fig. \ref{fig:gD_219}(b) compared
to those in Fig. \ref{fig:gD_219}(a) confirms the isomeric nature of the 25/2$^{+}$ state.
In addition to the above transitions, many new transitions were also observed in
the {\it delayed} gate of the 220-keV $\gamma$ ray as evident from the Fig. \ref{fig:gD_219}(b).
It was further observed that these new transitions were also present in all the
{\it early} coincidence spectra obtained with gates on the previously known transitions
following the decay of the 25/2$^{+}$ isomer, which unambiguously confirms their
assignment to $^{207}$At. The coincidence relationships of the new transitions feeding
the 2117-keV isomeric level and their placement in the level scheme are discussed in
Sec. \ref{sec:III}(A) in detail.

The half-life of the known 25/2$^{+}$ isomer at 2117 keV was also revisited 
using the decay-curve analysis method as shown in Fig. \ref{fig:halflife}.
This method is suitable to determine lifetimes significantly greater than the full width
at half-maximum (FWHM) of a prompt time distribution \cite{DecayCurv}.
For the present detector system, the FWHM was found to be around 50 ns for
the feeding and decaying $\gamma$-ray energies of around 500 keV. The deduced
half-life, $T_{1/2}$ = 107.5(9) ns, was found to be in good agreement with
the value reported earlier \cite{207At}.

Several new transitions were also identified below the 25/2$^{+}$ isomer.
For example, a 1055-keV transition was found to be in coincidence with the 220-keV
$\gamma$ ray as shown in Fig. \ref{fig:gD_219}(a). It was also observed to be
in coincidence with all the earlier known transitions above the 1085-keV level.
Sjoreen {\it et al.} \cite{207At} had established a 13/2$^{-}$ level at 1055 keV.
Also, the presence of an unobserved $\gamma$ ray of 30 keV between the 15/2$^-$
and 13/2$^-$ levels at 1085- and 1055-keV, respectively, was inferred on the basis
of $\gamma-\gamma$ coincidence relationships. The present study also supports these
observations and suggests a new direct decay path to the ground state via the 1055-keV
$\gamma$ ray.
%
%
\begin{figure}[t!]
\begin{center}
\includegraphics*[width=\columnwidth]{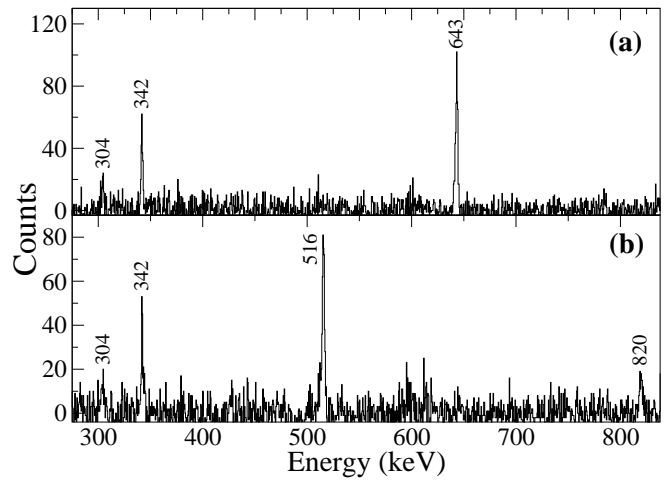}%
\caption{\label{fig_342} The coincidence spectra
with double gates on the (a) 472-, 516-keV, and (b) 472-, 643-keV transitions
to illustrate evidence of the new 304-, 342-, and 820-keV $\gamma$ rays.}
\end{center}
\end{figure}
%

\begin{figure}[t!]
\begin{center}
\includegraphics*[width=\columnwidth]{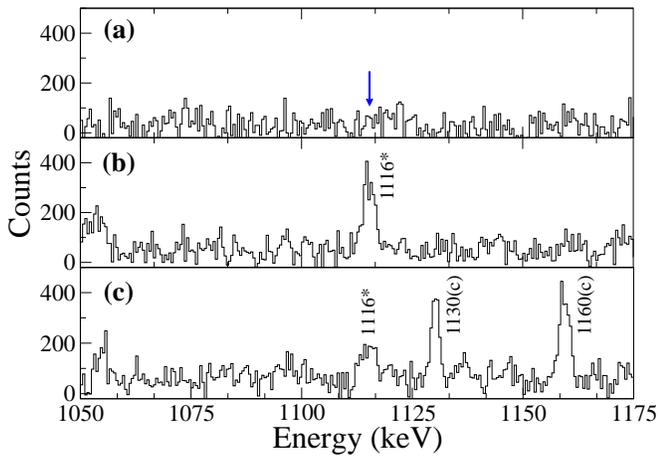}%
\caption{\label{fig:13by2} The $\gamma$-ray coincidence spectra in a gate on the
(a) 472-, (b) 516-, and (c) 342-keV $\gamma$-ray transitions illustrating the
absence/presence of the 1116-keV $\gamma$ ray in coincidence with the gating transitions.
The transitions labeled with (c) in the last panel are contaminants from
$^{63}$Cu and $^{67}$Ga produced through reactions of the beam with the Fe target frame.
}
\end{center}
\end{figure}

Furthermore, a sequence of 472-, 516-, and 342-keV $\gamma$-ray transitions
was established above the 11/2$^{-}$ level at 643 keV. It may be noticed that
Sjoreen {\it et al.} \cite{207At} had suggested a tentative 340-keV $\gamma$ ray
as opposed to the 342 keV one seen in this work. The $\gamma$-ray spectra
presented in Fig. \ref{fig_342} clearly establish the coincidence relationships of the new 342-keV transition
with the known 643-, 472-, and 516 keV $\gamma$ rays. These spectra also confirm the placement of the
new 304- and 820-keV $\gamma$-ray transitions as shown in Fig. \ref{LS}. Also, a new 1116-keV
$\gamma$-ray was found in coincidence with the 516- and 342-keV $\gamma$ rays, but not with
the 472-keV transition as illustrated in Fig. \ref{fig:13by2}. The 1116-keV
$\gamma$ ray de-excites the associated level directly to the ground state.
The spin assignment of $I$ = (13/2) and (15/2) to the 1116-, and 1631-keV levels,
respectively, is adopted from Ref. \cite{207At}. A spin assignment of $I$ = (17/2)
to the 1973-keV level as well as positive parity for the states under consideration
are proposed on the basis of similarities in their decay paths with those in the
lighter odd-$A$ At isotopes, and will be discussed in detail in Sec. \ref{sec:IV}(A).

The remaining level scheme below the 25/2$^{+}$ isomer was investigated using
similar $\gamma-\gamma$ coincidence relationships, relative intensity
considerations and the energy-sum technique. The procedure of the relative
intensity measurements is discussed later in this section.
The energy of the $\gamma$-ray transitions below the 25/2$^+$ isomer and their
relative $\gamma$-ray intensities are listed in Table \ref{table-I}.
The spin-parity assignments of the levels below the 25/2$^+$ isomer,
except for the 1115.5-, 1631.0-, and 1973.2-keV levels, are adopted from Ref. \cite{207At}.
%
%
%
\begin{figure*}[ht!]
 \includegraphics*[width=0.9\textwidth]{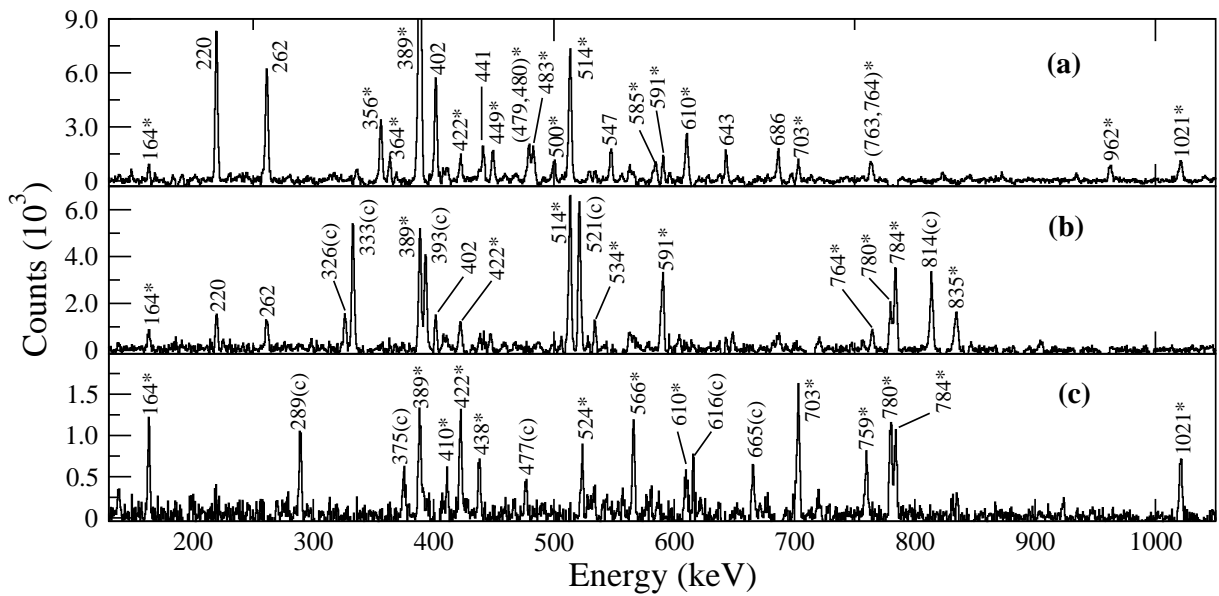}%
 \caption{\label{fig:g784_356_500} The $\gamma$-ray spectra illustrating the transitions
 in coincidence with the (a) 784-, (b) 356-, and (c) 500-keV $\gamma$ rays.
 The transitions marked with an asterisk are the new $\gamma $ rays assigned to $^{207}$At, while
 those marked with a symbol (c) are contaminants mainly from $^{196}$Pt, $^{206,207}$Po, and $^{206}$At.}
\end{figure*}
%
\begin{table}[t!]
\begin{ruledtabular}
\caption{\label{table-I} Table of $\gamma$-ray energies ($E_\gamma$), corresponding
level energies ($E_{i}$), relative $\gamma$-ray intensity ($I_\gamma$), and spin-parity of the
initial ($I_{i}^\pi$) and final ($I_{f}^\pi$) levels below the 25/2$^{+}$ isomer in $^{207}$At.
The relative intensities are quoted in percent by normalizing their values with respect to the
intensity of the 643-keV transition, which is assumed to be 100.
The listed spin-parities are adopted from Ref. \cite{207At}, except for the
1115.5-, 1631.0-, and 1973.2-keV levels. See the text for their spin-parity assignments.
The uncertainties in the $\gamma$-ray energies and relative intensities have contributions from both statistical
and systematic factors. The systematic uncertainty in $I_\gamma$ is considered to be 5$\%$ of the relative $\gamma$-ray intensity.}
\begin{center}

\begin{tabular}{ccccc}
$E_\gamma$ (keV)   & $E_{i}$ (keV) &  $I_\gamma$ &  $I_{i}^\pi$ & $I_{f}^\pi$ \\
\hline
(30)\footnotemark[1]	&1084.6(2)	 &$-$		&15/2$^-$	&13/2$^-$ \\
(43)\footnotemark[1]	&686.3(1)	 &$-$		&13/2$^-$	&11/2$^-$ \\
130.5(3)	&1495.1(2)	 &4.3(2)\footnotemark[2] 	&21/2$^{-}$ & $-$  \\
131.0(5)	&1364.6(5)	 &	$-$        & $-$           &17/2$^{-}$ \\
149.0(1)	&1233.6(1)	 &21.3(12) 	&17/2$^{-}$     &15/2$^{-}$ \\
219.6(1)	&2116.6(1)	 &53.8(31) 	&25/2$^{+}$     & 23/2$^{-}$ \\
261.5(1)	&1495.1(2)	 &134.7(77) &	21/2$^{-}$  & 17/2$^{-}$ \\
304.0(4)	&1935.3(5)	 &$<$1 	&$-$            & (15/2$^{+}$)\\
342.2(3)	&1973.2(3)	 &2.8(2) 	&(17/2$^{+}$)   & (15/2$^{+}$)\\
368.7(2)	&1054.8(4)	 &14.3(8) 	&13/2$^{-}$     & 13/2$^{-}$\\
398.2(2)	&1084.6(2)	 &16.9(10) 	&15/2$^{-}$     & 13/2$^{-}$\\
401.9(1)	&1897.0(1)	 &113.4(65) &	23/2$^{-}$  & 21/2$^{-}$ \\
412.0(2)	&1054.8(4)	 &13.0(7) 	&13/2$^{-}$     & 11/2$^{-}$ \\
441.4(1)	&1084.6(2)	 &60.5(34) 	&15/2$^{-}$     & 11/2$^{-}$ \\
472.3(2)	&1115.5(4)	 &6.7(4) 	&(13/2$^{+}$)   & 11/2$^{-}$ \\
515.5(3)	&1631.0(3)	 &3.7(2) 	&(15/2$^{+}$)   & (13/2$^{+}$) \\
547.3(1)	&1233.6(1)	 &56.1(32) 	&17/2$^{-}$     & 13/2$^{-}$ \\
643.2(1)	&643.2(1)	 &100.0(52) &	11/2$^{-}$  & 9/2$^{-}$ \\
678.3(2)	&1364.6(5)	 &9.7(6) 	&$-$            & 13/2$^{-}$ \\
686.3(1)	&686.3(1)	 &109.6(57) &	13/2$^{-}$  & 9/2$^{-}$ \\
722.4(3)	&1956.0(4)	 &1.7(1) 	&$-$            & 17/2$^{-}$ \\
819.8(3)	&1935.3(5)	 &1.2(1) 	&$-$            & (13/2$^{+}$) \\
871.4(2)	&1956.0(4)	 &5.2(3) 	&$-$            & 15/2$^{-}$ \\
1054.8(1)	&1054.8(4)	 &36.2(19) 	&13/2$^{-}$     & 9/2$^{-}$ \\
1115.5(4)   &1115.5(4)  &$<$1      &(13/2$^{+}$)   & 9/2$^{-}$ \\

\end{tabular}
\vspace*{-4mm}
\end{center}
\end{ruledtabular}
\footnotetext[1]{{Unobserved $\gamma$ ray, but its presence is inferred from the
coincidence relationships.}}
\footnotetext[2]{{Summed relative $\gamma$-ray intensity of the 130.5- and 131-keV $\gamma$ rays.}}
\end{table}
%
%
\subsection{Level structure above the 25/2$^{+}$ isomer}
As mentioned earlier, excited states in $^{207}$At were studied previously
by Sjoreen {\it et al.} up to 25/2$^{+}$ isomer \cite{207At}. However, no excited states
were known above this isomeric level. The present study extends the level scheme beyond
the 25/2$^{+}$ isomer up to 47/2$\hbar$ and $E_x$ $\approx$ 6.5 MeV. As illustrated in
Fig. \ref{fig:gD_219}(b), many new transitions were identified which precede the
220-keV $\gamma$ ray. Four new sequences consisting of about 30 new $\gamma$-ray transitions
were established, which feed the known 25/2$^{+}$ isomeric state.

The {\it early} $\gamma$-ray spectra obtained with gates on the $\gamma$ rays following
the decay of the 25/2$^+$ isomer indicate that the 784- and 389-keV transitions are the
strongest transitions above the isomer. These transitions were also found to be
in mutual coincidence [see Fig. \ref{fig:g784_356_500}(a)], which justifies their
placement above the 2117-keV isomeric level as shown in Fig. \ref{LS}.
A cascade of new transitions viz. 514-, 356-, and 591 keV, was found to be
in coincidence with the above transitions.
These transitions were placed above the 3289-keV level in sequence ``A''
in the descending order of their relative intensities as shown in Fig. \ref{LS}.
Another cascade of 1021-, 703-, 500-, 422-, 438-, and 164-keV transitions,
labeled as sequence ``D'' in Fig. \ref{LS}, was also identified to be in
coincidence with the 784- and 389-keV $\gamma$ rays. The prompt coincidence
spectra illustrated in Fig. \ref{fig:g784_356_500} support the coincidence
relationships of these transitions. Furthermore, the new 483-, 479-, 449-, 962-, and
364-keV transitions were also observed in coincidence with the 389- and 784-keV $\gamma$ rays.
These transitions were placed in sequence ``B'' as shown in Fig. \ref{LS}
by using their relative intensities and coincidence relationships.

The coincidence relationships and placements of the new 780-, 759-, 566-, and 524-keV
transitions observed in Fig. \ref{fig:g784_356_500}(c) are discussed in the next
section.

\begin{figure}[b!]
\begin{center}
\includegraphics*[width=\columnwidth]{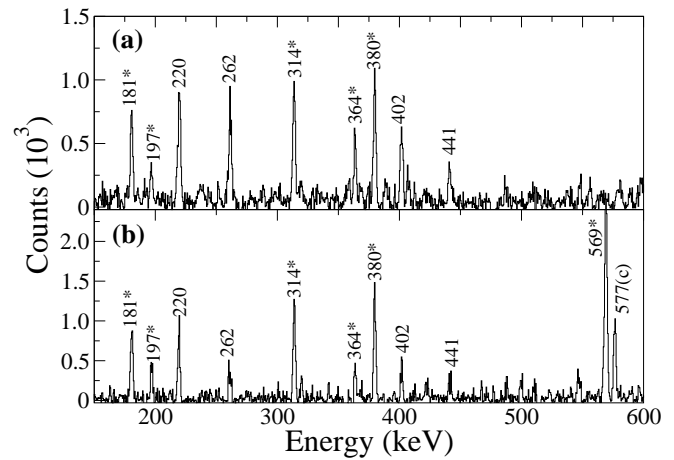}%
\caption{\label{fig:seqC} The $\gamma$ ray coincidence spectra obtained with gate on the
(a) 1192- and (b) 623-keV $\gamma$-ray transitions. The transitions marked with an asterisk
are the new transitions  from $^{207}$At. The 577-keV transition marked with symbol (c) is
contamination from $^{207}$Po.
}
\end{center}
\end{figure}

Several other new transitions, viz., the 569-, 1192-, 623-, 181-, 380-, and 314 keV,
were also observed in {\it early} coincidence with the 220-keV
[see Fig. \ref{fig:gD_219}(b)] as well as with the other transitions following
the decay of the 25/2$^{+}$ isomer. These transitions were not observed in
coincidence with the transitions in sequences ``A'', ``B'', and ``D'', except
the 364-keV one (see Fig. \ref{fig:g784_356_500}). Two illustrative prompt spectra with
gates on the 1192- and 623-keV transitions are presented in Fig. \ref{fig:seqC}.
The observed new transitions were placed in sequence ``C'', which directly feeds
the 25/2$^{+}$ isomer.

Furthermore, some transitions were identified in the present work which bypass
the 25/2$^{+}$ isomer. Two transitions, viz. the 780- and 1230-keV ones were observed
in mutual coincidence and placed above the 21/2$^{-}$ level at 1495 keV on the
basis of their relative intensities and coincidence relationships. In addition,
the presence of new 197- and 364-keV transitions and their placement in the level
scheme connect the sequence ``C'' with the one of the bypassing transitions.
Also, it was noticed that the 364-keV $\gamma$ ray is a doublet and both
the components are in coincidence with each other. The above coincidence relationship and
coincidence of the 364-keV $\gamma$ ray with all the transitions
in sequence ``C'' suggest the presence of an unobserved 69-keV transition between the
4252- and 4183-keV levels. However, a possibility of the presence of an another 364-keV
transition which may directly feed the 4183-keV level can not be discarded.
%
%
\begin{table*}[ht!]
\begin{ruledtabular}
\caption{\label{table-II} Table of $\gamma$-ray energies ($E_\gamma$), corresponding initial level energies ($E_{i}$),
spin-parity of the initial ($I_{i}^\pi$) and final ($I_{f}^\pi$) levels, relative $\gamma$-ray intensity
($I_\gamma$), $R_\textrm{DCO}$, $\Delta_\textrm{asym}$ values and assigned multipolarity of the $\gamma$-ray
transitions de-exciting the levels above the 25/2$^+$ isomer.
The relative intensities are quoted in percent by normalizing their values with respect to the
intensity of the 643-keV transition, which is assumed to be 100.
The uncertainties in the $\gamma$-ray energies, relative
intensities and $R_\textrm{DCO}$ values have contributions from both statistical and systematic factors.
The systematic uncertainty in $I_\gamma$ is considered to be 5$\%$ of the relative $\gamma$-ray intensity.}
\begin{center}

\begin{tabular}{cccccccc}
 $E_\gamma$ (keV)   & $E_{i}$ (keV) &  $I_{i}^\pi$ & $I_{f}^\pi$  &  $I_\gamma$  & $R_\textrm{DCO}$ & $\Delta_\textrm{asym}$ &Multipolarity\\
\hline
(69)\footnotemark[1]  &4251.8(4)   &35/2$^-$       &33/2$^{(+)}$  &$-$       &                          &              &\\
(70)\footnotemark[1]  &3289.3(2)   &33/2$^{+}$     &31/2$^{+}$    & $-$      &                          &              &\\
124.7(3)   &3289.3(2)   &33/2$^+$       &33/2$^+$      &1.4(1)    &                          &              &\\
163.6(2) 	&6538.4(3)	  &47/2~~ 	      &45/2~~	         &3.7(2)	&0.52(4)\footnotemark[2]   & 	          &$D$\\
180.6(2)	&3489.4(3)	  &29/2$^{(+)}$	  &27/2$^{-}$	 &4.5(3)	&1.24(9)\footnotemark[3]   &		      &$D$\\
193.1(3)	&2149.1(4)	  & $-$ 	      &$-$           &3.8(2)    &                          &              &\\
196.8(3)	&3505.3(5)	  &29/2$^{(-)}$	  &27/2$^{-}$	 &3.7(2)    &                          & 	          &\\
252.0(3)	&2149.1(4)	  & $-$           & 23/2$^{-}$   &1.1(1)    &                          &              &  \\
259.1(4)	&4159.2(5)	  &37/2$^+$	      &35/2$^+$      &$<$1      &                          & 		     &\\
264.7(2)	&3952.8(3)	  &37/2~~     	  &35/2$^+$	     &3.8(2)	&0.46(3)\footnotemark[2]   &		     &$D$\\
313.7(2)	&4182.9(3)	  &33/2$^{(+)}$	  &31/2$^{(+)}$	 &9.7(5)	&1.01(7)\footnotemark[3]   &-0.112(29)   &	$M$1\\
356.4(2)	&4159.2(5)	  &37/2$^+$	      &35/2$^+$	     &9.8(6)	&0.48(3)\footnotemark[2]   &-0.087(21)   &$M$1\\
363.6(2)	&4615.4(3)	  &37/2$^{(-)}$	   &35/2$^-$	     &5.4(3)	&0.50(4)\footnotemark[2]   &-0.025(35)   &($M$1 + $E$2)\\
363.8(5)	&3869.2(5)	  &31/2$^{(+)}$	  &29/2$^{(-)}$  &7.5(6)      &                          &		     &\\
379.8(2)	&3869.2(5)	  &31/2$^{(+)}$	  &29/2$^{(+)}$	 &5.2(3)	&0.94(7)\footnotemark[3]   &-0.139(33)   &$M$1\\
388.6(1)	&3289.3(2)	  &33/2$^+$	      &29/2$^+$	     &36.3(20)	&1.07(8)\footnotemark[2]   &0.101(18)    &$E$2\\
410.5(4)	&4310.3(4)	  &37/2$^+$	      &35/2$^+$	    &$<$1       &                          &		     &\\
422.5(2)	&5936.6(3)	  &43/2$^{(+)}$ &41/2$^+$	    & 6.3(4)	&0.54(4)\footnotemark[2]   &-0.016(15)   &($M$1 + $E$2)\\
438.2(3)	&6374.8(3)	  &45/2~~	      &43/2$^{(+)}$	    &3.4(2)	    &0.60(5)\footnotemark[2]   &		     &$D$\\
449.4(2)	&4251.8(4)	  &35/2$^-$	      &35/2$^+$	    &5.6(3)	    &1.17(9)\footnotemark[2]&-0.077(26)   &$E$1,$\Delta I$ = 0\\
479.3(3)	&4251.8(4)	  &35/2$^-$	      &33/2~~     	&1.5(1)	    &1.21(9)\footnotemark[3]   &		     &$D$\\
479.6(3)	&3380.3(4)	  &	  $-$         & 29/2$^+$    & 2.4(2)    &                          &		     &\\
483.2(2)	&3772.5(2)	  &33/2~~     	  &33/2$^+$	    &3.5(2)	    &0.86(6)\footnotemark[2]   &		     &$D$, $\Delta I$ = 0\\
487.6(1)	&2384.6(1)	  &29/2$^{+}$     & 23/2$^{-}$  &66.5(38)   &                          &             &\\
500.4(2)	&5514.1(5)	  &41/2$^+$	      &39/2$^+$	    &6.1(4)	    &0.44(3)\footnotemark[2]   &-0.045(21)   &$M$1\\
513.5(1)	&3802.8(2)	  &35/2$^+$	      &33/2$^+$	    &22.1(13)	&0.48(3)\footnotemark[2]   &-0.066(18)   &$M$1\\
523.5(2)	&3688.1(3)	  &35/2$^+$	      &33/2$^+$	    &14.3(8)	&1.02(8)\footnotemark[3]   &-0.044(21)   &$M$1\\
534.4(3)	&4693.6(3)	  &39/2~~	          &37/2$^+$	    &2.0(1)	    &1.03(8)\footnotemark[3]   &		     &$D$\\
566.3(2)	&4254.4(3)	  &37/2$^+$	      &35/2$^+$	    &6.8(4)	    &1.10(8)\footnotemark[3]   &-0.054(21)   &$M$1\\
569.3(1)	&2685.9(2)	  &27/2$^+$	      &25/2$^+$	    &15.1(8)	&0.86(6)\footnotemark[3]   &-0.091(32)   &$M$1\\
584.8(3)	&3485.5(4)	  &$-$	          &29/2$^+$     &1.7(1)     &                          &		     &\\
590.6(2)	&4749.8(3)	  &39/2~~	          &37/2$^+$	    &4.8(3)	    & 0.92(7)\footnotemark[3]  &	         &$D$\\
610.5(2)	&3899.8(3)	  &35/2$^+$	      &33/2$^+$	    &8.5(5)	    & 0.44(3)\footnotemark[2]  &-0.072(23)   &$M$1\\
622.9(2)	&3308.8(4)	  &27/2$^{-}$	  &27/2$^+$	    &4.1(2)	    & 2.22(16)\footnotemark[3] &		     &$D$, $\Delta I$ = 0\\
654.0(3)	&2149.1(4)	  &$-$            & 21/2$^{-}$  &2.4(1)     &                          &             &  	 \\
703.4(2)	&5013.7(4)	  &39/2$^+$	      &37/2$^+$	    &3.7(2)	    & 0.50(4)\footnotemark[2]  &	-0.040(28)	     &$M$1\\
759.4(3)	&5013.7(4)	  &39/2$^+$	      &37/2$^+$	    &2.3(1)	    & 0.95(7)\footnotemark[3]  &-0.027(51)   &($M$1 + $E$2)\\
763.2(3)	&3663.9(4)	  &$-$ 	          &29/2$^+$ 	&2.5(2)     &                          &		     &\\
764.4(3)	&5514.1(5)	  &41/2$^+$	      &39/2~~     	&1.9(1)	    &0.98(7)\footnotemark[3]   &		     &$D$\\
779.8(2)	&2274.9(3)	  &25/2$^{-}$	  &21/2$^-$     &9.6(5)	    &1.97(15)\footnotemark[3]  &	 0.069(23)        & $E$2  \\
780.0(1)    &3164.6(1)	  &33/2$^+$	      &29/2$^+$   	&42.0(24)	& 2.14(16)\footnotemark[3] &0.076(25)    &$E$2\\
784.1(1)	&2900.7(1)	  &29/2$^+$	      &25/2$^+$   	&54.4(30)	&0.94(7)\footnotemark[2]   &0.098(11)	&$E$2\\
834.6(1)	&3219.2(1)	  &	31/2$^+$      & 29/2$^+$    &31.0(18)   &0.81(6)\footnotemark[3]   &-0.029(25)	&$M$1\\
854.0(2)	&2349.1(3)	  &$-$            &21/2$^{-}$               &3.5(2)                    &             &  	\\
934.5(3)	&4223.8(4)	  &35/2~~	          &33/2$^+$     &2.6(2)	   &0.44(3)\footnotemark[2]    &	    	&$D$\\
947.6(3)	&4900.4(4)	  & $-$	          &37/2~~         &1.9(1)    &                           &	     	&\\
962.5(2)	&4251.8(4)	  &35/2$^-$	      &33/2$^+$   	&4.3(3)	   &0.53(4)\footnotemark[2]	   &0.188(33)	&$E$1\\
1021.0(2)	&4310.3(4)	  &37/2$^+$	      &33/2$^+$  	&5.7(3)	   &1.09(8)\footnotemark[2]	   &0.047(38)	&$E$2\\
1071.9(3)	&4361.2(4)	  &	 $-$          & 33/2$^+$    &1.9(1)    &                           &	    	&\\
1191.9(2)	&3308.8(4)	  &27/2$^{-}$	  &25/2$^+$  	&6.4(4)	   & 1.11(8)\footnotemark[3]   &0.155(56)	&$E$1\\
1230.4(3)	&3505.3(5)	  &29/2$^{(-)}$	  &25/2$^{-}$ &2.1(1)	   &1.11(8)\footnotemark[2]	   &	    	&$Q$\\
1354.7(3)	&5514.1(5)	  &41/2$^{+}$	  &37/2$^{+}$   &2.8(2)	   &                           &	    	& \\
\end{tabular}
\vspace*{-4mm}
\end{center}
\end{ruledtabular}
\footnotetext[1]{{Unobserved $\gamma$ ray, but its presence is inferred from the
coincidence relationships.}}
\footnotetext[2]{{$R_\textrm{DCO}$ value is obtained with a gate on a stretched quadrupole transition.}}
\footnotetext[3]{{$R_\textrm{DCO}$ value is obtained with a gate on a pure dipole transition.}}
\end{table*}
%
%

The spin-parity of the new levels above the 25/2$^{+}$ isomer were assigned
on the basis of measured $R_\textrm{DCO}$ and $\Delta_\textrm{asym}$ values
for the corresponding feeding and/or depopulating $\gamma$-ray transitions.
As mentioned in Sec. \ref{sec:II}, the $R_\textrm{DCO}$ value was found to be
$\approx$ 1.0 when the gating transition and the $\gamma$ ray of interest have the
same multipolarity, while $R_\textrm{DCO}$ was obtained to be $\approx$ 2.0 (0.5)
for a stretched quadrupole (dipole) transition with a gate on a pure dipole
(stretched quadrupole) one. These values were estimated from a few $\gamma$ rays
of known multipolarity from other nuclei populated in the same reaction. In $^{207}$At,
the states above the 25/2$^{+}$ isomer are established for the first time in the
present study. However, the isomeric nature of the 25/2$^{+}$ state [$T_{1/2}$ = 107.5(9) ns]
prevents the use of the transitions below the 25/2$^{+}$ isomer to obtain the $R_\textrm{DCO}$
values for the new transitions feeding the isomer.
Therefore, the $R_\textrm{DCO}$ values for a few new transitions of the sequence ``A''
viz. 784-, 389-, 514-, and 356 keV, were determined using mutual gates and their
multipolarities were assigned on the basis of consistency in their $R_\textrm{DCO}$
values as discussed below. The $R_\textrm{DCO}$ values for the 389-, 514-, and 356-keV
transitions with a gate on the 784-keV $\gamma$ ray are 1.07(8), 0.48(3), and 0.48(3),
respectively. These values suggest that both the 784- and 389-keV $\gamma$-ray transitions
are of the same multipolarity. Also, these $R_\textrm{DCO}$ values collectively suggest that
the 784- and 389-keV transitions are $\Delta I$ = 2 transitions, while the 514- and
356-keV transitions are of $\Delta I$ = 1 multipolarity. Furthermore, the $R_\textrm{DCO}$
values for the 784-, 389-, and 514-keV transitions were determined using a gate on the
356-keV $\gamma$ ray and found to be 2.00(14), 2.14(15), and 0.94(7), respectively.
These values further corroborate the preceding multipolarity assignments of the 784-,
389-, 514- and 356-keV transitions. The $R_\textrm{DCO}$ values for the remaining transitions
in sequences ``A$-$D'' were obtained using the deduced multipolarities of the
784-, 389-, 514- and 356-keV transitions. Further, electric or magnetic nature of the
new transitions were determined from the measured $\Delta_\textrm{asym}$ values.
The information on the $\gamma$-ray energies, relative intensities, $R_\textrm{DCO}$
and $\Delta_\textrm{asym}$ values for the transitions above the isomeric 25/2$^{+}$
level are listed in Table \ref{table-II}.

\subsection{Evidence of a 29/2$^+$ isomer at $E_x$ = 2385 keV}
%
%
\begin{figure}[t!]
\begin{center}
\includegraphics*[width=\columnwidth]{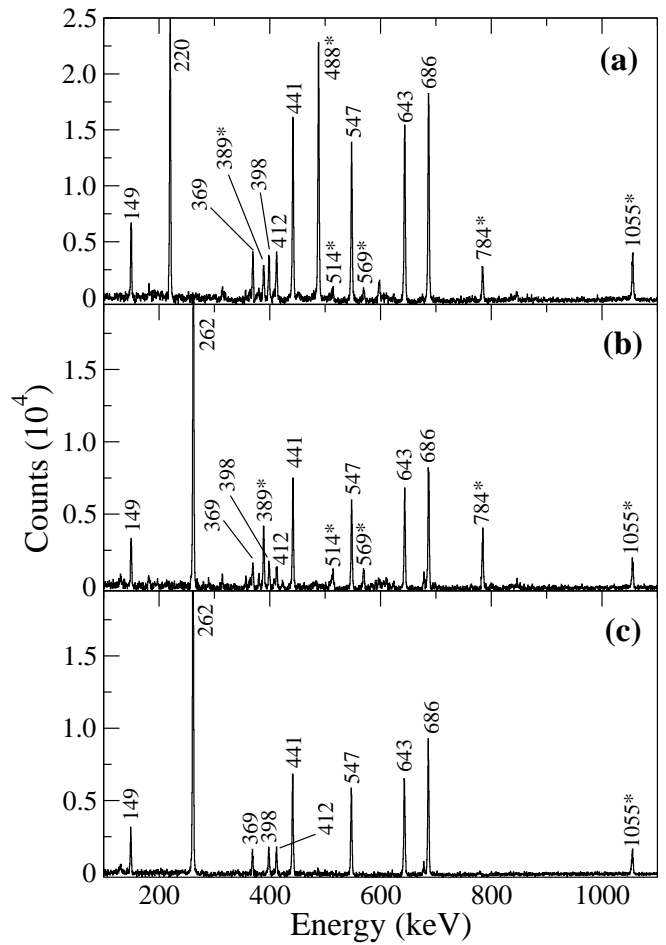}%
\caption{\label{fig:29by2} Double-gated coincidence spectra illustrating 
$\gamma$-ray transitions in gates on the (a) 402-, 262-keV, (b) 402-, 220-keV,
and (c) 402-, 488-keV $\gamma$ rays within the prompt coincidence window.
The new transitions are marked with asterisks.
}
\end{center}
\end{figure}
%
An isomeric 29/2$^{+}$ state is known in $^{199,201,203,205,209,211}$At nuclei
\cite{201At,203At,205At_1,205At_2,209At_1,209At_2,209At_2,211At}. However, such
a state was not known in $^{207}$At \cite{207At}. In this section, we discuss the
first evidence of the 29/2$^{+}$ isomeric state in $^{207}$At.
Figure \ref{fig:29by2} illustrates the double-gated coincidence spectra of
the (a) 402-, 262 keV, and (b) 402-, 220-keV transitions. It was observed
that a strong 488-keV transition is present in Fig. \ref{fig:29by2}(a),
but not in Fig. \ref{fig:29by2}(b). This indicates that the new 488-keV $\gamma$
ray is in coincidence with the 262- and 402-keV transitions but not with the 220 keV,
which in turn suggests that the 488-keV transition should be placed above the 402-keV
$\gamma$ ray. The placement of the 488-keV $\gamma$ ray introduces a new level at 2385 keV.
Further, Fig. \ref{fig:29by2}(c) depicts the $\gamma$-ray transitions in a double gate
of the 402- and 488-keV $\gamma$ rays. It may be noted that only transitions below the
1495-keV level (see Fig. \ref{LS}) are observed in Fig. \ref{fig:29by2}(c), which further
corroborates the above placement of the new 488-keV $\gamma$ ray. The absence of any
prompt $\gamma$ ray/s feeding the 2385-keV level indicates the presence of a long-lived
isomeric state. Here, two possibilities may be considered. One could be that the 2385-keV
level itself has a large half-life and therefore, the transition/s feeding the 2385-keV
level could not be observed in the prompt coincidence (100 ns) with the 488-keV transition.
Another possibility could be the presence of a closely-spaced higher-lying
metastable state, which may de-excite to the 2385-keV level via one or more unobserved
low energy and/or high multipole transitions. However, the second possibility can be
discarded on the basis of the coincidence relationships of the transitions which are
observed in {\it early} coincidence with the new 488-keV $\gamma$ ray as discussed below.

\begin{figure}[t!]
\begin{center}
\includegraphics*[width=\columnwidth]{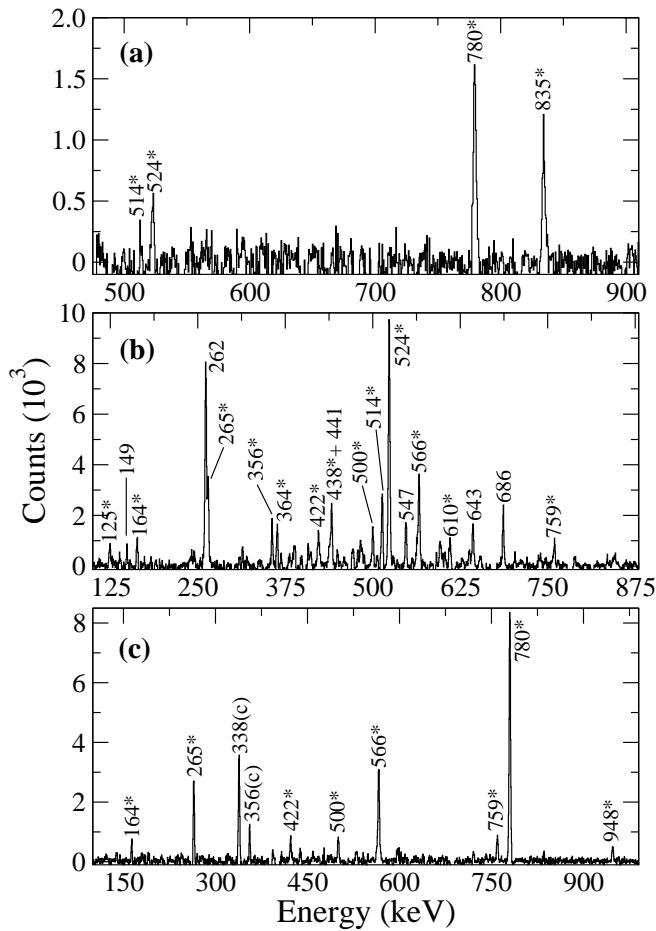}%
\caption{\label{fig:seqE} Coincidence spectra illustrating $\gamma$-ray transitions
in coincidence with the (a) 488-keV $\gamma$ ray within a 100$-$275 ns {\it early} time window,
(b) 780-, and (c) 524-keV $\gamma$ rays within the prompt coincidence window. The new transitions
from $^{207}$At are marked with asterisks. The transitions marked with the symbol (c) are
contaminants from $^{206}$Bi and $^{196}$Pt.}
\end{center}
\end{figure}

Figure \ref{fig:seqE}(a) illustrates the transitions which precede the
488-keV $\gamma$ ray within the 100$-$275 ns coincidence window. It was further
observed that the strongest 780- and 835-keV $\gamma$-ray transitions are not
in mutual coincidence, while the 780-keV one is in coincidence with a weaker 524-keV
transition as evident from the two lower panels of Fig. \ref{fig:seqE}.
The new 780-, 524-, 265-, and 566-keV transitions were placed above the 2385-keV
level as shown in Fig. \ref{LS} on the basis of their relative intensities
and $\gamma-\gamma$ coincidence relationships. In addition, a new 759-keV transition
was also observed in coincidence with the 780-, 524-, and 566-keV $\gamma$ rays.
The placement of the 759-keV $\gamma$ ray connects the sequences ``D'' and ``E''
and also corroborates the placement of the new 780-keV transition.

%
%
%
Furthermore, the $\gamma$-ray transitions feeding the 33/2$^+$ level at 3289 keV
in sequence ``A'' are also found in coincidence with the 780- and 835-keV
transitions, but not with the 524-keV one
[see Figs. \ref{fig:seqE}(b), \ref{fig:seqE}(c), and \ref{fig:g784_356_500}(b)].
It may be noted that the presence of the 262-, 441-, 547-, 643-, and
686-keV $\gamma$-ray transitions in Fig. \ref{fig:seqE}(b) is attributed
to their coincidence with another 780-keV transition which is placed
above the 1495-keV level, as already mentioned in Sec. A. Also, a
new 125-keV $\gamma$ ray is found to be in coincidence with the 780-keV
transition and with those feeding the 33/2$^{+}$ level at 3289 keV.
The placement of the new 125-keV $\gamma$ ray connects the sequences
``A'' and ``E'', and hence further validates the positions of the new
levels and transitions above the 2385-keV state as shown in Fig. \ref{LS}.
Moreover, the coincidences of the 835-keV line with the transitions feeding
the 3289-keV level indicate the presence of an unobserved 70-keV transition.
The placement of the transitions above the isomers in the sequences ``A'',
``D'', and ``E'' along with that of the 488-keV transition uniquely establishes
a state at 2385 keV and the fact that it has a long half-life.

The spin-parity of the 2385-keV level is inferred based on the following considerations.
The $I^{\pi}$ = 39/2$^{+}$ for the 5014-keV level is firmly established on the basis of the $R_\textrm{DCO}$
and $\Delta_\textrm{asym}$ values of the 703-, 1021-, 389- and 784-keV transitions.
Now, this state is connected to the 2385-keV level via three $\Delta I$ = 1 transitions,
viz., 759-, 566-, and 523-keV and one $\Delta I$ = 2 (780 keV) transition (see Table II).
These measurements suggest $I$ = 29/2 for the 2385-keV level, which in turn
leads to a possible $E$3 or $M$3 multipolarity for the 488-keV transition. Furthermore,
intensity balance at the 1897-keV level (assuming $I$ = 29/2 for the 2385-keV level,
as discussed above) supports $E3$ multipolarity for the 488-keV transition,
and hence $I^{\pi}$ = 29/2$^{+}$ is proposed for the 2385-keV state.
Furthermore, the $\Delta_\textrm{asym}$ values for the 780-, 523-, and 566-keV transitions
suggest positive parity of the 3165-, 3688-, and 4254-keV levels in sequence ``E''.
The proposed spin-parity, $I^\pi$ = 29/2$^{+}$, for the isomeric 2385-keV level is also supported by the
energy systematics of the known 29/2$^{+}$ isomeric states in the neighboring odd-$A$
At isotopes and will be discussed in Sec. \ref{sec:IV}(B) in detail.

An attempt to determine the half-life of the isomeric state under consideration
using the electronic timing method resulted in an almost flat time-difference distribution curve of the
depopulating and feeding transitions.
This suggests a significantly longer half-life as compared to the
available narrow time range viz., 0$-$275 ns.
However, a rough estimate of the half-life can be obtained as follows.
The delayed gates of the 220- and 488-keV transitions were used to
determine the intensities of the 784- and 780-keV transitions within
the 100$-$275 ns {\it early} coincidence window, respectively.
The ratio of the intensities of the 784- to 780-keV transitions was found to be 13.8(10).
Now, the amount of the initial intensities (See Table. \ref{table-II}) of the 784- and 780-keV
transitions which decay within the 100$-$275 ns were determined with the help of
radioactive decay law using the known half-life [107.5(9) ns] of the 25/2$^{+}$ state
and assuming various values of $T_{1/2}$ for the 29/2$^{+}$ state.
The ratio of these intensities is to be
compared with the experimental value mentioned above.
Among the various values assumed for the half-life of the 29/2$^{+}$ state,
it was observed that $T_{1/2} \approx$ 3.5 $\mu$s
is able to reproduce the experimental ratio 13.8(10).
In addition, a possible range of the half-life (2$-$4.5 $\mu$s) is surmised on
the basis of a systematic study of the $B(E3)$ rates for the
29/2$^{+}\rightarrow$ 23/2$^{-}$ transitions in the neighboring odd-A At
isotopes and will be discussed in \ref{sec:IV}(B).
Finally, we note that the intensity ratios obtained using the radioactive decay
law for the limits of expected half-life from the $B(E3)$ systematics viz., 2 and 4.5 $\mu$s are
around 8.1 and 17.6, respectively.
\subsection{Intensity measurements}
The relative intensities of the $\gamma$-ray transitions listed in Tables \ref{table-I}
and \ref{table-II} were determined using the data at 87 MeV.
In order to obtain the relative
intensities of the transitions below the isomers, the pure 643- and 441-keV transitions
were fitted in a spectrum obtained from the single-fold data. Further, the deduced
intensities were normalized with respect to the intensity of the 643-keV $\gamma$ ray,
for which the intensity of 100 units was assumed. The relative $\gamma$-ray intensities
of the 686-, 1055-, and 1116-keV transitions were determined by measuring the branching
ratios of all the transitions feeding directly to the ground state in the efficiency
corrected summed spectrum with the gates on the relevant transitions. Further, the
efficiency corrected sum gate of all the transitions feeding the ground state
was used to determine the intensities of the remaining transitions below the isomers.
The intensities of the transitions below the isomers were normalized using the relative
intensity of the 441-keV $\gamma$ ray.

The relative intensities of the $\gamma$-ray 
transitions above the isomers were determined as follows.
As discussed in the earlier sections, the 780-keV transition
is a doublet; (i) 25/2$^{-}$ $\rightarrow$ 21/2$^{-}$ transition feeding the
21/2$^{-}$ level at 1495 keV, and (ii) 33/2$^{+}$ $\rightarrow$ 29/2$^{+}$ transition
which feeds directly to the 29/2$^{+}$ isomer at 2385 keV. The summed intensity of
the 780-keV peak was obtained from single-fold data and normalized with respect
to the relative intensity of the 643-keV transition.
Further, an individual relative intensity of the bypassing 780-keV
[25/2$^{-}$ $\rightarrow$ 21/2$^{-}$] transition was deduced in a gate of the
262-keV $\gamma$ ray. It may be noted that the second 780-keV transition, which
populates the 29/2$^{+}$ isomer, was not observed in the prompt coincidence (100 ns)
with the transitions following the decay of the 29/2$^{+}$ isomer
(see Fig. \ref{fig:29by2}). Thus, the intensity obtained with the gate on the 262-keV
transition corresponds only to the relative intensity of the
25/2$^{-}$ $\rightarrow$ 21/2$^{-}$, 780-keV transition. The relative intensity
of the second 780-keV transition could then be determined by subtracting the contribution
of the relative intensity of the bypassing 780-keV transition from the summed
relative intensity.
Further, the intensity of the 835-keV transition above the 29/2$^{+}$
isomer was determined from the ratio of the intensities of the 780-keV and
835-keV transitions in a {\it delayed} gate of the 488-keV $\gamma$ ray
[see Fig. \ref{fig:seqE}(a)]. The intensity of the 784-keV $\gamma$ ray,
relative to the total $\gamma$-ray intensity of the  643-keV transition,
was also determined from the single-fold data. In order to deduce the relative
intensity of the 389-keV $\gamma$-ray, the 389- and 784-keV peaks were fitted
in a prompt coincidence spectrum with a gate on the 220-keV $\gamma$ ray.
The deduced intensities of both the transitions were normalized with respect 
to the relative intensity of the 784-keV transition which was obtained from 
single fold-data as discussed above. 
The relative intensities of the remaining transitions in the sequences ``A'',
``B'', ``D'', and ``E'' were determined in efficiency corrected summed gates
of the 780-, 784-, and 835-keV transitions and normalized with respect to
the relative intensity of the 389-keV $\gamma$ ray.
The relative intensity of the 364-keV [37/2$^{(-)}$ $\rightarrow$ 35/2$^{-}$]
transition in sequence ``B'' is corrected for the branch which
de-excites the 4252-keV level in sequence ``C''.
Moreover, the intensity of the second 364-keV $\gamma$ ray, which populates the 3505-keV
level was determined by subtracting the contribution of the upper 364-keV
[37/2$^{(-)}$ $\rightarrow$ 35/2$^{-}$] transition from the total relative intensity
of the 364-keV doublet. The total relative intensity of the doublet was determined in a
efficiency corrected summed prompt spectrum obtained with gates on the 780-, 784-, 835-,
1192-, and 569-keV transitions.

Finally, the relative intensities of the 1192-, 569-, 623-, 181-, 380, and 197-keV
transitions of sequence ``C'' were also deduced by fitting these transitions in a
gate of the 220-keV $\gamma$ ray. The relative intensity of the 784-keV transition (obtained from single-fold data)
was used for normalization to deduce the relative intensities of the above transitions.
\section{DISCUSSION}\label{sec:IV}
Nuclear structure investigations in the region around the doubly closed-shell
nucleus $^{208}$Pb are of immense interest as a variety of nuclear structure
aspects have been observed in this region. In particular, the level structure
in At isotopes with a few neutron-holes with respect to the $N$ = 126 shell
closure is governed by single-particle excitations
\cite{211At, 209At_1,209At_2,209At_3,209At_4}.
Furthermore, shears bands have been reported in the lighter At isotopes
viz., $^{201,203,204}$At \cite{201At,203At,204At}.
A similar sequence of $\Delta I$ = 1 transitions had also been reported
in $^{205}$At \cite{205At_2}. However, the shears structure of the observed
sequence was not confirmed \cite{205At_2}. On the other hand, the lighter Po and At
nuclei are known to exhibit oblate and prolate deformed shapes \cite{192-195Po,197At}.
Thus, a systematic study across the isotopic chain enhances the understanding
of shape evolution with decreasing neutron number.
%
\begin{figure}[t!]
\begin{center}
\includegraphics*[width=\columnwidth]{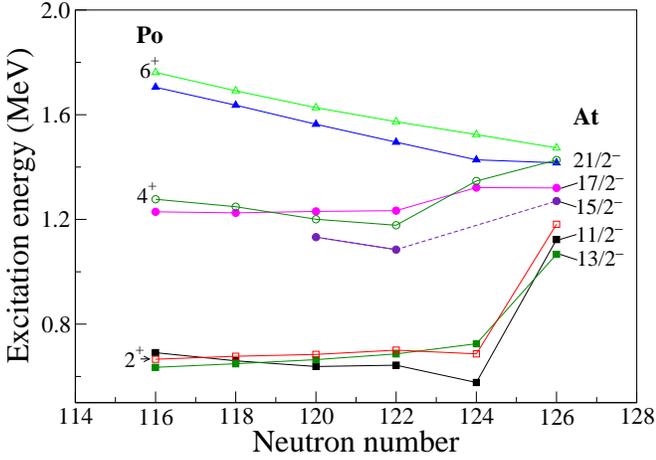}%
\caption{\label{fig:systematics}Energy systematics of selected low-lying states in At
(solid symbols) and Po (open symbols) isotopes.
The excitation energies of the states in At and Po isotopes are taken from present work and Refs.
\cite{201At,203At,205At_1,209At_4,211At,200_202Po,204Po,208Po,210Po}.
}
\end{center}
\end{figure}
%
In $^{207}$At, which is the subject of the present work, the excited
states only up to the 25/2$^{+}$ isomer at 2117 keV were known \cite{207At}.
The present study reports an extended level scheme of $^{207}$At.
In addition to several new transitions below the isomer, four new sequences
are identified which feed the known 25/2$^{+}$ isomer.
Moreover, the new 29/2$^{+}$ isomer is also established at 2385 keV,
which completes the systematics of this isomer in odd-$A$ At isotopes.
The proposed level scheme is understood in the framework of the shell-model
approach. The yrast and near-yrast states in $^{207}$At are also compared with the
systematics of the corresponding states in Po and odd-$A$ At isotopes for
a qualitative understanding as discussed in the following sections.

Large-scale shell model calculations have been performed with the KHM3Y effective
interaction \cite{208Pb_brown}, which was developed for the $Z$ = 50$-$126, $N$ = 82$-$184 model space.
The model space comprised of 24 orbitals
in total; the proton orbitals include $g_{7/2}$, $d_{5/2}$, $h_{11/2}$, $d_{3/2}$,
$s_{1/2}$ below the $Z=82$ and $h_{9/2}$, $f_{7/2}$, $i_{13/2}$, $f_{5/2}$,
$p_{3/2}$, $p_{1/2}$ above it, while the neutron orbitals involve $i_{13/2}$, $p_{3/2}$,
$f_{5/2}$, $p_{1/2}$, $h_{9/2}$, $f_{7/2}$ below the $N=126$ and
$g_{9/2}$, $i_{11/2}$, $j_{15/2}$, $g_{7/2}$, $d_{5/2}$, $d_{3/2}$, $s_{1/2}$ above. 
The cross-shell two-body matrix elements (TBMEs) are based on the M3Y interaction \cite{m3y}.
The neutron-proton particle-particle and hole-hole TBMEs use the Kuo-Herling  interaction \cite{kuo}
as modified in Ref. \cite{mod_kuo}. In order to make the calculation feasible,
the proton orbitals below the $Z$ = 82 shell closure were completely filled and the
neutron excitations across the $N$ = 126 shell were not allowed.
The calculations were performed using the shell model code NuShellX \cite{NuShellX}.
Recently, the experimental results of $^{207}$Tl \cite{207Tl, 207Tl_Emma, 208Pb} and $^{208}$Pb
\cite{208Pb_brown,208Pb} were interpreted using the shell model calculations
which also used the same interaction mentioned above.
%
\begin{figure}[t!]
\begin{center}
\includegraphics*[width=0.9\columnwidth]{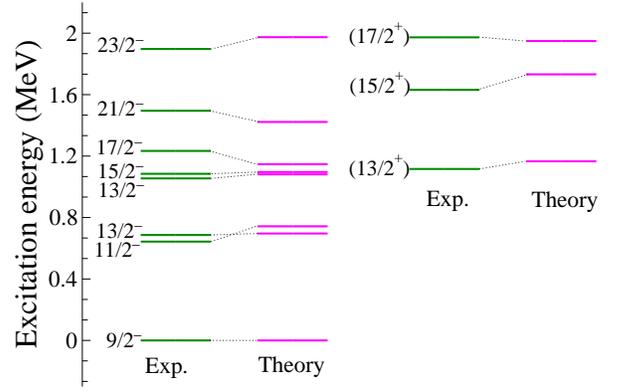}%
\caption{\label{fig:SM_belowisomers} Comparison of the experimental level energies of the yrast and
near-yrast states below the isomers with those predicated from the shell model calculations.}
\end{center}
\end{figure}
%
%
\begin{figure*}[t!]
\begin{centering}
  \includegraphics[angle = 270, width=\textwidth]{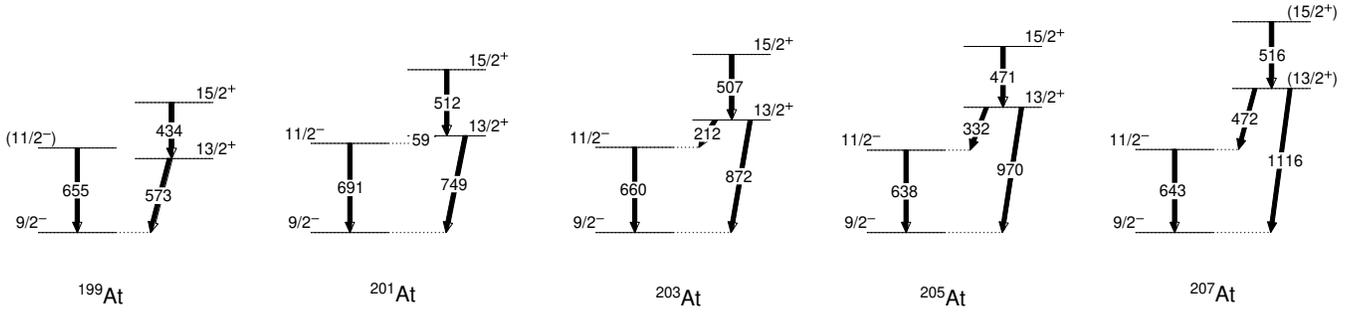}%
\end{centering}
\caption{\label{13by2} Evolution of the 13/2$^{+}$ state in odd-$A$ At isotopes with increasing
neutron number between 114 $\leq N \leq$ 122. The experimental information of the presented levels
of the $^{199-207}$At nuclei are taken from the present work and
Refs. \cite{201At,203At, 201_203At, 205At_1, 205At_2, 207At}.
}
\end{figure*}
\subsection{States below the isomers}
Figure \ref{fig:systematics} depicts the energy systematics of the
yrast negative-parity states, viz. 11/2$^{-}$, 13/2$^{-}$, 15/2$^-$,
17/2$^{-}$, and 21/2$^{-}$, in odd-$A$ At isotopes, which is compared
with the 2$^{+}$, 4$^{+}$, and 6$^{+}$ states of the corresponding Po isotones.
It is evident from the figure that the (11/2$^{-}$, 13/2$^{-}$), (15/2$^-$,
17/2$^{-}$), and 21/2$^{-}$ states in odd-$A$ At isotopes closely follow the
trend in excitation energies of the 2$^{+}$, 4$^{+}$, and 6$^{+}$ states
in the corresponding Po isotones, respectively.
The observed similarity in the excitation energies of the yrast states
suggests that the low-lying negative parity states in odd-$A$ At isotopes
can be understood in terms of the coupling of an unpaired proton in
$h_{9/2}$ orbital to the respective states of the even-even Po core.
A similar interpretation is noted to explain the low-lying
negative-parity states in neighboring odd-$A$ At isotopes
\cite{201At, 203At, 205At_1,205At_2, 199At}.
Also, a sudden increase in the excitation energy of the
11/2$^-$ and 13/2$^-$ [2$^{+}$($^{210}$Po)$\otimes \pi h_{9/2}$]
states may be noted, as opposed to the same states in the lighter
isotopes where excitation energies are found to be rather constant.
The marked difference in the excitation energies of the yrast states
in the $N$ = 126 isotones relative to those in the lighter isotopes
can be attributed to the difference in their configurations.
The low-lying yrast states in the closed neutron shell nuclei,
viz. $^{210}$Po and $^{211}$At are governed by pure proton configurations,
while both proton-particle and neutron-hole excitations are expected to
play a role in the lighter Po and At isotopes. The yrast states in
$^{211}$At ($N$ = 126) are well understood in terms of the
$\pi(h_{9/2}^{3})$, $\pi(h_{9/2}^{2}f_{7/2})$,
and $\pi(h_{9/2}^{2}i_{13/2})$ configurations \cite{211At}.

Figure \ref{fig:SM_belowisomers} shows a comparison of the experimental
level energies of the yrast and near-yrast states below the isomers with
those predicted from the shell-model calculations.
It is evident from the figure that the experimental level energies are
in good agreement (within 100 keV) with the shell-model
predictions. The low-lying negative parity states viz. 11/2$^{-}$,
13/2$^{-}$, 15/2$^-$, 17/2$^{-}$, and 21/2$^{-}$ are predicted to originate
from the $\pi (h_{9/2}^3$)$\otimes$  $\nu (f_{5/2}^{-2}p_{1/2}^{-2})$
configuration.
The energy systematics of these states in Po and At isotopes also suggests
the same configuration as discussed above. The primary configuration
(with 25.6\% parentage) for the 23/2$^{-}$ state is predicated to be
$\pi(h_{9/2}^{2}f_{7/2})$ $\otimes$  $\nu (f_{5/2}^{-2}p_{1/2}^{-2})$.
It may be noted that only the states with firm spin-parity assignments
are compared with the shell model calculations, except the (13/2$^{+}$),
(15/2$^{+}$), and (17/2$^{+}$) states. The positive parity of these states
is suggested from the similarity in the level structures in odd-A At isotopes
as discussed below.

The 13/2$^{+}$ state, which originates from the $\pi$($i_{13/2}$) configuration,
has been identified in several odd-$A$ At isotopes
\cite{197At,199At,201At,203At, 201_203At, 205At_1, 205At_2,211At}.
Figure \ref{13by2} illustrates the evolution of the 13/2$^{+}$ state with
increasing neutron number in At isotopes with 114 $\leq N \leq$ 122.
It is observed that the excitation energy of the 13/2$^+$ state
increases with increasing number of neutrons. However, the excitation energy
of the 11/2$^-$ state remains nearly constant as shown in Fig. \ref{13by2}.
A similar trend in the excitation energy of the 13/2$^{+}$ state has also been
reported in odd-$A$ Bi isotopes \cite{203_205Bi,199_201Bi,199Po_betadecay}.
This feature is explained by the coupling of the unpaired proton in the $i_{13/2}$
orbital to the increasing number of valence neutron holes.
The interaction of the $\pi \nu ^{-1}$ configuration for both the particles and
holes in the $i_{13/2}$ orbital has the smallest repulsion, which brings the
13/2$^{+}$ state down in energy with a decreasing number of neutrons \cite{199At,195Bi}.
In  $^{199}$At, the 13/2$^{+}$ state deexcites to the ground state only
via an $M$2 branch unlike the heavier isotopes, viz. $^{201,203,205}$At,
where it is observed to decay via $E$1 and $M$2 transitions to
the 11/2$^{-}$ and 9/2$^{-}$ ground state, respectively.
A similar structure is also observed in $^{207}$At, where the 1116-keV
state de-excites via the 472- and 1116-keV $\gamma$ rays to the 11/2$^{-}$
level at 643 keV and the ground state, respectively. The observed
similarities in the excitation energy and the decay paths of the 1116-
and 1631-keV levels  in $^{207}$At with the kindred states in the
neighboring At isotopes suggest that these states are also likely to
have positive parity. As illustrated in Fig. \ref{fig:SM_belowisomers},
the excitation energies of the proposed positive parity states
viz. (13/2$^{+}$), (15/2$^{+}$), and (17/2$^{+}$), are in very good
agreement with the shell model calculations, which suggest a dominant
$\pi(h_{9/2}^2i_{13/2})\otimes \nu (f_{5/2}^{-2}p_{1/2}^{-2})$
configuration for the states under consideration.

\subsection{Isomers and their decay properties}\label{secIV:C}
The region around the doubly-magic nucleus $^{208}$Pb is rich in nuclear
isomerism \cite{AKJ}. The presence of high-$j$ valence orbitals leads to
the realization of isomeric states in this region which arise due to
dominant intrinsic degrees of freedom. An isomer (38$^+$) in $^{212}$Rn ($N$ = 126)
at $E_x$ = 12.5 MeV is the highest-lying isomer known to date \cite{212Rn}.
This isomer results from the triple-core excitations coupled with
the aligned valence protons. Several high-spin isomeric states have
been reported in Po, At, and Fr isotopes \cite{AKJ}.
A 25/2$^{+}$ isomer is reported in $^{203,205,207}$At isotopes
\cite{203At,205At_1,205At_2,207At}.
Also, a long-lived 29/2$^{+}$ state is known in all odd-$A$ At isotopes
with 114 $\leq N \leq$ 126, except in $^{207}$At
\cite{201At, 203At, 205At_1, 205At_2, 207At, 209At_1,209At_2, 209At_3, 209At_4,211At}.
As discussed in the Sec. \ref{sec:III}, we have identified the
long-lived 29/2$^{+}$ isomeric state in $^{207}$At at $E_x$ = 2385 keV.

\begin{figure}[t!]
\begin{center}
\includegraphics*[width=\columnwidth]{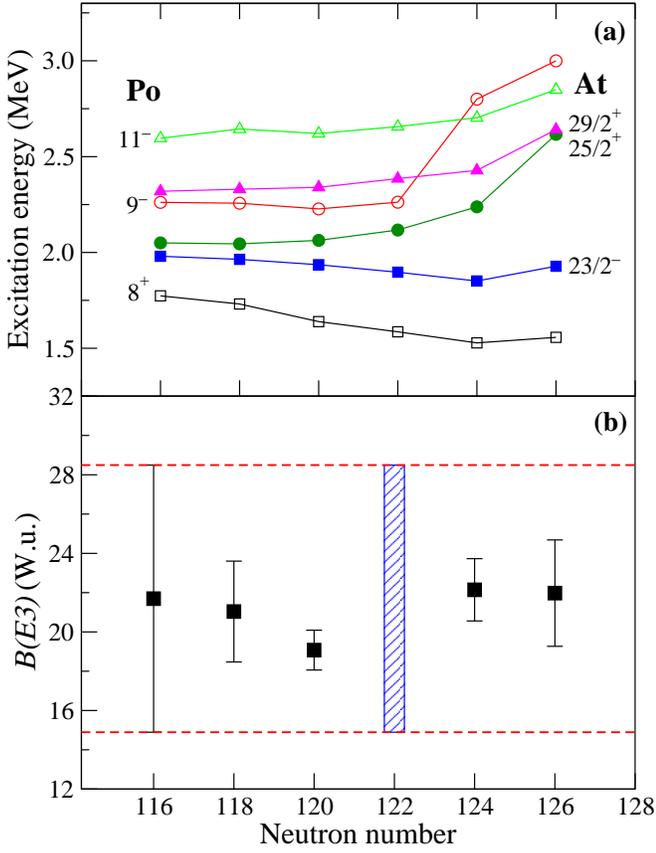}%
\caption{\label{fig:isomers}(a) Energy systematics of the 29/2$^{+}$, 25/2$^{+}$ isomers
and the 23/2$^{-}$ state in odd-$A$ At isotopes (solid symbols) and the
11$^{-}$, 9$^-$ and 8$^{+}$ states of the  corresponding Po (open symbols) isotones.
Panel (b) of the figure presents the variation of the experimental $B$({\it E}3) values
for the 29/2$^+$ $\rightarrow$ 23/2$^{-}$ transition in odd-$A$ At isotopes with
increasing neutron number. The shaded area indicates the possible range of the
$B$({\it E}3) value for the 488-keV transition in $^{207}$At.
The experimental information for the At and Po isotopes is taken from the present work and Refs.
\cite{NDS_A201, NDS_A203, NDS_A205, NDS_A209, NDS_A211, 200_202Po,204Po,208Po,210Po}.
}
\end{center}
\end{figure}

The high-spin isomers in odd-$A$ At isotopes may be interpreted in terms of
the coupling of an unpaired proton in specific orbitals to the even-even core
of the corresponding Po isotones. The 11$^{-}$ and 9$^-$ isomeric states have
been reported in even-even Po isotopes, which originate from the
$\pi(h_{9/2}^{1}i_{13/2}^{1})_{11^{-}}$ two-quasi-proton and
$\nu(f_{5/2}^{-1}i_{13/2}^{-1})_{9^{-}}$ two-neutron-hole configurations,
respectively \cite{198_200Po,202_204_206Po,208Po,210Po}.
Figure \ref{fig:isomers}(a) illustrates the energy systematics of the
11$^{-}$, 9$^{-}$, and the subsequent 8$^{+}$ states in the even-even
Po isotopes, which is compared with the corresponding 29/2$^{+}$,
25/2$^{+}$, and 23/2$^{-}$ levels in the odd-A At isotones.
The systematic trend in the excitation energies suggests that
the structure of the 29/2$^{+}$ and 25/2$^{+}$ isomeric states 
in $^{207}$At may be interpreted in terms of a weak coupling of an
unpaired proton in the $h_{9/2}$ orbital to the 11$^{-}$ and 9$^{-}$
isomers in $^{206}$Po, respectively. Similarly,
the 23/2$^{-}$ state may be realized from a weak coupling of the
unpaired proton in the $f_{7/2}$ orbital to the 8$^{+}$ state of the Po core.
Similar considerations have been used to explain the origin of
the isomeric states in other odd-$A$ At isotopes
\cite{199At, 201_203At, 201At, 203At, 205At_1, 205At_2, 209At_1, 209At_2,209At_3,209At_4,211At}.
%
%

The 25/2$^{+}$ isomer in $^{207}$At de-excites to the 23/2$^{-}$ state
via the 220-keV $E$1 transition. The experimental $B(E1)$ value for the
220-keV transition is 3.6(9)$\times$10$^{-7}$ $e^2fm^2$, which is hindered
by seven orders of magnitude with respect to the single-particle estimate.
The hindrance in the decay of the 25/2$^{+}$ state may be attributed to a
considerable difference in the wave functions of the 25/2$^{+}$ and 23/2$^{-}$
states. In $^{203,205}$At, the 25/2$^{+}$ isomer de-excites via competing $E$1
and $E$2 transitions unlike the one in $^{207}$At where only an $E$1 decay branch
is observed \cite{203At,205At_1,205At_2}.

The systematics of the excitation energy and the decay path for the newly
identified 29/2$^{+}$ isomer is observed to account well for its proposed spin-parity
when compared to the same state in the neighboring odd-$A$ At isotopes.
In the lighter At isotopes, viz.  $^{201,203, 205}$At, the 29/2$^{+}$ isomer
is known to de-excite via $E$2 and $E$3 transitions to the subsequent 25/2$^{+}$
and 23/2$^{-}$ levels, respectively \cite{201At, 203At, 205At_1, 205At_2},
while it decays only via an $E$2 branch in $^{199}$At \cite{199At}.
It has been observed that the $E$2 decay-branch becomes weaker with
increasing neutron number. The 29/2$^{+}$ isomer in $^{209}$At
is known to decay only via an $E$3 transition \cite{209At_1, 209At_2,209At_3,209At_4}.
However, in the case of $^{211}$At this isomer is reported to deexcite via
an unobserved low-energy $E$2 transition, in addition to a strong $E$3 decay path \cite{211At, 211At_E2}.
The proposed 29/2$^{+}$ isomeric level in $^{207}$At is also observed to decay via
the 488-keV $E$3 transition. However, the present data do not provide any conclusive evidence for
an $E$2 decay branch.

As mentioned earlier, the direct measurement of the half-life of the 29/2$^+$
isomer was not possible. However, the systematics of the $E$3 transition strengths in odd-$A$ At
isotopes can be used to determine the range of the expected half-life.
Figure \ref{fig:isomers}(b) depicts the variation of the $B(E3)$ values
corresponding to the 29/2$^+$ $\rightarrow$ 23/2$^{-}$ transitions in odd-$A$
At isotopes with increasing neutron number. It may be noticed that the plotted
$B(E3)$ values are roughly constant.
The observed striking similarities in the properties of the 29/2$^+$ isomer
in odd-$A$ At isotopes as discussed above suggest that the transition strength
of the 488-keV (29/2$^+$ $\rightarrow$ 23/2$^{-}$) $E$3 transition in $^{207}$At
can assume a value between the range displayed by the shaded area in
Fig. \ref{fig:isomers}(b). This range of $B(E3)$ values in turn suggests
a half-life between 2$-$4.5 $\mu$s for the 29/2$^+$ isomer in $^{207}$At.
It is worth mentioning that the rough estimate of the half-life ($\approx$ 3.5 $\mu$s) for
the 29/2$^+$ state as outlined in the previous section is consistent with the above range.

The shell-model calculations suggest that the 29/2$^{+}$ and 23/2$^-$
levels are three quasi-proton states with $\pi(h_{9/2}^2i_{13/2}^1)$ and $\pi(h_{9/2}^{2}f_{7/2}^1)$
primary configurations, respectively. The 25/2$^{+}$ isomer is associated with the
$\pi(h_{9/2}^3)_{9/2^{-}} \otimes \nu(f_{5/2}^{-1}i_{13/2}^{-1})_{9^{-}}$
configuration and originates from a weak coupling of the $\pi h_{9/2}$
with two-quasi-neutron-hole configuration. Further, it was observed that
the shell-model calculations account well for excitation energies of the
23/2$^-$ and 25/2$^{+}$ states, while an underestimation by 295 keV is
observed for the 29/2$^{+}$ isomer. It may be noted that the 29/2$^+$
isomer decays via the 488-keV $E$3 transition, which corresponds to the
$\pi (i_{13/2} \rightarrow f_{7/2}$) transition as suggested by the shell-model
configurations for the states under consideration. In such cases, where the initial
and final state wave-functions involve the high-$j$ proton ($i_{13/2}$ and $f_{7/2}$)
and neutron ($j_{15/2}$ and $g_{9/2}$) orbitals which differ by $\Delta j = \Delta l = 3$,
the configuration of initial state is likely to have a mixing of the 3$^-$ octupole
phonon of the $^{208}$Pb coupled with the final state \cite{fast_E3}.
Therefore, the corresponding $E$3 transitions are expected to be very
fast \cite{fast_E3}.
Such states have been observed in a number of nuclei around the
doubly-closed $^{208}$Pb \cite{208Pb,207Tl, E3_transitions}.
The 29/2$^{+}$ isomeric level in $^{207}$At also appears to have a
mixing of the octupole vibrational phonon. A recent study by Berry {\it et al.}
reported that the states involving octupole vibrations in the region around $^{208}$Pb
are underestimated ($\approx$ 250 keV) by the shell-model calculations performed using the
KHM3Y effective interaction \cite{207Tl}. A similar effect is also observed in the
calculations for the 29/2$^{+}$ state in $^{207}$At.

\subsection{States above the isomers}
\begin{figure}[t!]
\begin{center}
\includegraphics*[width=0.9\columnwidth]{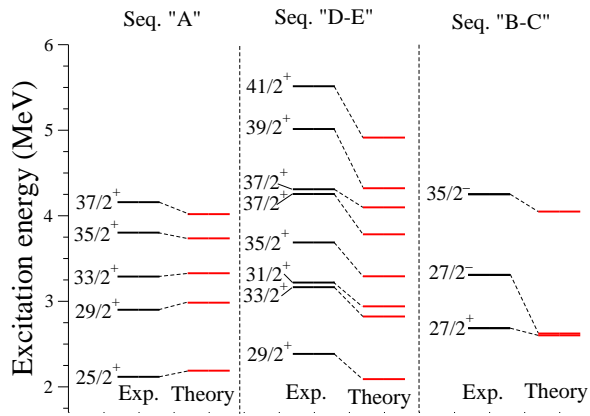}%
\caption{\label{fig:SM_aboveisomers} Comparison of the experimental level energies of yrast and
near-yrast states above the isomers with that predicted by the shell model calculations.
Only those states for which firm spin-parities are known are considered for the comparison.}
\end{center}
\end{figure}

Figure \ref{fig:SM_aboveisomers} shows a comparison of the excitation
energies of the yrast and near-yrast states above the 25/2$^{+}$ isomer
with the shell-model calculations. The shell-model states corresponding
to the experimental levels in a specific sequence are chosen on the basis of configurations.
As discussed in the previous section, the isomeric 25/2$^{+}$ state is associated with the
$\pi (h_{9/2}^3)_{9/2^-}\otimes \nu(f_{5/2}^{-1}i_{13/2}^{-1})_{9^-}$
configuration. The same configuration is observed to account well (within 150 keV),
as shown in Fig. \ref{fig:SM_aboveisomers},
for the higher-lying states in sequence ``A'' by breaking of the proton pair in the $h_{9/2}$ orbital.
A similar sequence has also been reported in $^{205}$At, which feeds the 25/2$^{+}$
isomer at 2063 keV \cite{205At_2}. However, no such sequence was observed in $^{203}$At \cite{203At}.
%

As mentioned earlier, the 29/2$^{+}$ isomeric level is predominantly the three quasi-proton state
with a maximum possible angular momentum from the $\pi(h_{9/2}^2i_{13/2})$ 
configuration. For the realization of the states above the 29/2$^{+}$ isomer,
the neutron-hole pairs may contribute in the angular-momentum generation.
Therefore, the shell-model states originating from the $\pi(h_{9/2}^2i_{13/2})_{29/2^+}\otimes I_n $
configuration, where $I_n$ is the contribution to the angular momentum from the neutron holes
are compared with the states in sequences ``D'' and ``E''.
As shown in Fig. \ref{fig:SM_aboveisomers}, a systematic underestimation is observed in the
calculated excitation energies of the states in sequences ``D'' and ``E''.
It was further noticed that the deviation between the experiment and the shell-model
calculations increases with excitation energy, which may be attributed to the
truncation scheme adopted for the calculations, due to computational limitations.
The states in sequences ``B'' and ``C'' are found to be in rather good agreement with the
shell-model calculations, except the 27/2$^{-}$ state.

Moreover, the available spectroscopic information on the states above the 29/2$^{+}$ isomer
is scarce in the neighboring odd-$A$ At isotopes for a systematic comparison.

\section{SUMMARY}
Excited states above the known 25/2$^{+}$ isomer
in $^{207}$At have been investigated for the first time using the
$^{198}$Pt($^{14}$N, 5{\it n}$\gamma$)$^{207}$At
reaction. The half-life of the 25/2$^+$ isomer is revisited and a value
of $T_{1/2}$ = 107.5(9) ns is inferred, which is consistent with that
reported by Sjoreen {\it et al.} \cite{207At}. The level scheme is
considerably extended up to 6.5 MeV and 47/2$\hbar$ with the addition
of about 60 new $\gamma$-ray transitions. A long-lived 29/2$^{+}$
isomeric state ($E_x$ = 2385 keV) is established for the first time
in $^{207}$At, with a similar level having been identified earlier
in the neighboring odd-$A$ At isotopes.
The half-life of the proposed 29/2$^{+}$
isomer is estimated to be in the 2$-$4.5 $\mu$s range on the basis of
the systematics of the $B(E3)$ values corresponding to the $E$3 transitions,
which de-excite the 29/2$^{+}$ isomer in the neighboring odd-$A$ At isotopes.
Furthermore, the 1116-keV level is suggested to originate from
the $\pi (i_{13/2})$ configuration on the basis of similarities
in its excitation energy and decay paths with that of the 13/2$^{^+}$
state in the neighboring odd-$A$ At isotopes. The large-scale
shell-model calculations were performed using the shell-model code NuShellX
with the KHM3Y effective interaction.
An overall good agreement has been observed between the experimental results
and the shell-model predictions for the states below the isomers and those
in sequence ``A'' above the 25/2$^{+}$ isomer. However, the shell-model calculations
appear to significantly underestimate the excitation energies of the remaining states.
In addition, the weak coupling of an unpaired proton in the $h_{9/2}$ or $f_{7/2}$
orbitals with the states of $^{206}$Po core is observed to qualitatively account
for the level structure in $^{207}$At. An experiment to determine half-life of the
29/2$^+$ isomer is in order.

\section{ACKNOWLEDGMENTS}
The authors would like to thank the technical staff from IUAC for their support
during the experiment. The financial support by DST, India (Grant No. IR/S2/PF-03/2003-III)
for the INGA project is also acknowledged. K. Y. acknowledges the financial assistance from
MHRD, India. A. Y. D. and D. S. would like to acknowledge the financial support by SERB (DST),
Grant No. CRG/2020/002169. Madhu acknowledges financial support from DST, India under the
INSPIRE fellowship scheme (IF 180082). P.C.S. acknowledges a research grant from SERB (India),
CRG/2019/000556.

\end{document}